%% file: main.tex
\documentclass[sigconf]{acmart}

\usepackage{xcolor}
\usepackage{booktabs}
\usepackage{longtable}
\newif\ifdraft
\drafttrue

\input{macros}

\AtBeginDocument{%
  \providecommand\BibTeX{{%
    \normalfont B\kern-0.5em{\scshape i\kern-0.25em b}\kern-0.8em\TeX}}}

\setcopyright{acmcopyright}
\copyrightyear{2023}
\acmYear{2023}
\acmDOI{XXXXXXX.XXXXXXX}

\acmConference[MSR'23]{Make sure to enter the correct
  conference title from your rights confirmation emai}{May 15--16,
  2023}{Melbourne, Australia}
\acmPrice{15.00}
\acmISBN{978-1-4503-XXXX-X/18/06}

\begin{document}

\title{Public Sector Open Source Software Projects - How is development organized?}


\author{Johan Linåker}
\affiliation{%
 \institution{RISE Research Institutes of Sweden}
 \country{Sweden}}
\email{johan.linaker@ri.se}

\author{Björn Lundell}
\affiliation{%
 \institution{University of Skövde}
 \country{Sweden}}
\email{bjorn.lundell@his.se}

\author{Francisco Servant}
\affiliation{%
 \institution{Universidad Rey Juan Carlos}
 \country{Spain}}
\email{francisco.servant@urjc.es}

\author{Jonas Gamalielsson}
\affiliation{%
 \institution{University of Skövde}
 \country{Sweden}}
\email{jonas.gamalielsson@his.se}

\author{Sachiko Muto}
\affiliation{%
 \institution{RISE Research Institutes of Sweden}
 \country{Sweden}}
\email{sachiko.muto@ri.se}

\author{Gregorio Robles}
\affiliation{%
 \institution{Universidad Rey Juan Carlos}
 \country{Spain}}
\email{grex@gsyc.urjc.es}


\renewcommand{\shortauthors}{Linåker et al.}

\begin{abstract}
    \textit{Background/Context}: Open Source Software (OSS) started as an effort of communities of volunteers, but its practices have been adopted far beyond these initial scenarios. For instance, the strategic use of OSS in industry is constantly growing nowadays in different verticals, including energy, automotive, and health. For the public sector, however, the adoption has lagged behind even if benefits particularly salient in the public sector context such as improved interoperability, transparency, and digital sovereignty have been pointed out. When Public Sector Organisations (PSOs) seek to engage with OSS, this introduces challenges as they often lack the necessary technical capabilities, while also being bound and influenced by regulations and practices for public procurement.

    \textit{Objective/Aim:}
    We aim to shed light on how public sector OSS projects, i.e., projects initiated, developed and governed by public sector organizations, are developed and structured. We conjecture, based on the challenges of PSOs, that the way development is organized in these type of projects to a large extent disalign with the commonly adopted bazaar model (popularized by Eric Raymond~\cite{raymond1999cathedral}), which implies that development is carried out collaboratively in a larger community. 
    
    \textit{Method:}
    We plan to compare and contrast public sector OSS projects with a set of earlier reported case studies of bazaar OSS projects, including Mockus et al.'s reporting of the Apache web server and Mozilla browser OSS projects~\cite{mockus2002two}, along with the replications performed on the FreeBSD, JBossAS, JOnAS, and Apache Geronimo OSS projects~\cite{dinh2005freebsd, ma2010developers}. To enable comparable results, we will replicate the methodology used by Mockus et al. on a purposefully sampled subset of public sector OSS projects. The subset will be identified and characterized quantitatively by mining relevant software repositories, and qualitatively investigated through interviews with individuals from involved organizations.
    

\end{abstract}



\maketitle

\section{Introduction}

Open Source Software (OSS) has since its inception, more than twenty years ago (although present as a phenomenon even longer), evolved from a community approach to scratching itches to becoming an established way for the industry to collaborate on common functionality~\cite{robles2019twenty}. Across industry sectors, competitors are pooling resources and leveraging communities to solve common problems, allowing innovation to be focused at the top of the stack. The result is that OSS is now present in a large majority of companies' internal code bases, as well as in products and services, and in our common digital infrastructure at large~\cite{tidelift2018survey, synopsis2020survey, osi2022state}. While Public Sector Organizations (PSOs) could also leverage OSS projects and collaboration to make better use of resources in developing government services, OSS uptake in the public sector has not gained the momentum that many anticipated~\cite{kovacs2004open}. Even though many administrations at the local, regional, national, and even supra-national levels have carried out a large amount of initiatives to develop their software projects as OSS and benefit from its advantages, the impact of such initiatives has so far been moderate.

In this study, we are interested in exploring this context, and what we refer to as public sector OSS projects, i.e., those that are \textit{initiated, developed, and governed by one (or multiple) PSOs, either through internal or commissioned resources}. Specifically, we aim to explore how the development and social structures of this type of projects compare to the more extensively explored community and industry-driven types of OSS projects~\cite{capra2008framework}.

A common means for characterizing the social structure of OSS projects is through the so-called ``onion''~\cite{crowston2005social} model: the main contributors are the core of the project (the ``core'' developers), \emph{surrounded} by a new layer of occasional contributors. In an outer layer, we can find the \emph{end users}. The further from the center, the less input and influence an individual has on the project. In large OSS projects, the outer layers of the ``onion'' are orders of magnitude superior in terms of individuals compared to the inner ones. This periphery of individuals commonly consists of users and casual contributors to the OSS project.

Projects that manage to grow an active and vibrant community commonly function as a \emph{bazaar}, as proposed in Eric Raymond's well-known \emph{The Cathedral and the Bazaar} essay~\cite{raymond1999cathedral}, with \emph{light} development and communication processes, \emph{fluid} management, and open to everybody to contribute. The \emph{cathedral}, in contrast, is a model that follows the \emph{traditional} setting commonly found within commercial development organizations (and in other OSS projects as well), which contextualized in an OSS project implies that planning and development are concentrated in one or a few individuals in isolation from the community~\cite{capiluppi2007cathedral}. 

Scholars have noted that a specific OSS project's development model, over time, may change its way of working from a development model characterised as a cathedral model to bazaar model~\cite{capiluppi2007cathedral}. Moreover, the same scholars claim that the development models for an OSS project which manages to gather attention and external contributors typically transition from a cathedral model to a bazaar model. In this regard, we consider the onion model as only being applicable and relevant to OSS projects in, or on the transition towards, the bazaar model.

In this study, our overarching research goal is to investigate how public sector OSS projects are developed and organized, and by extension whether the results may be characterized through the bazaar or cathedral development models~\cite{raymond1999cathedral, capiluppi2007cathedral}, or if new metaphors may be needed. We conjecture that \emph{after more than 20 years of initiatives around OSS in the public sector, development in public sector OSS projects to a large extent are organized in ways dis-aligning with the commonly-adopted bazaar model}. By extension, we assume that the extent to which the onion model applies is fundamentally different for public sector OSS projects~\cite{nakakoji2002evolution} compared to bazaar OSS projects~\cite{raymond1999cathedral} like those originally investigated by Mockus et al.~\cite{mockus2002two}. 

This has implications, as these projects cannot take advantage of bazaar-related benefits that come with a large developer community. Another implication may be that research and practice for community and industry-driven projects do not apply to public sector OSS projects. If this is the case, other approaches (managerial, technical, and processes-related) may have to be explored and evaluated. To the best of our knowledge, we are not aware of any scientific effort that has explored this conjecture. 

We propose to address this gap by comparing and contrasting the characteristics of public sector OSS projects with a set of previously reported OSS projects where development was carried out in line with the bazaar model (from now on referred to as bazaar OSS projects). We specifically look at the seminal work by Mockus et al. who studied the Apache web server~\cite{mockus2000case} and Mozilla browser OSS projects~\cite{mockus2002two}. Their study was first replicated (and mostly confirmed) on the community-driven FreeBSD OSS project~\cite{dinh2005freebsd}, and later on in the industry-driven JBossAS, JOnAS, and Apache Geronimo OSS projects~\cite{ma2010developers}.
To enable comparable results, we will replicate the methodology, and a relevant subset of research questions from the original study by Mockus et al~\cite{mockus2002two}.



Next, we provide further contextualization to this study. Following, we describe in detail our hypotheses and research questions, and their underpinning rationale. This is followed by the research design for how we aim to execute the study overall and address each research question in detail.

\section{Background and related work}


It has been reported that the scale of Europe's institutional capacity related to OSS is disproportionately smaller than the scale of the value created by OSS~\cite{blind2021impact}. Several factors contribute to this circumstance, including the potential for economic growth, innovation, and competition~\cite{greenstein2014digital, nagle2019government, blind2021impact}. OSS has also been shown to bring benefits that are particularly salient in the public sector context, among them improved interoperability~\cite{lundell2017addressing}, transparency~\cite{ofe2022strategic}, and digital sovereignty~\cite{nagle2022strengthening}.

The adoption, on the other hand, is dampened by challenges such as lacking technical capabilities and competency regarding software development~\cite{borg2018digitalization}, and issues related to public procurement regulations and procurement practices which impact on conditions for how OSS can be procured by PSOs~\cite{lundell2021enabling}. General knowledge of OSS is another concern, both within the PSOs and its vendors, e.g., in terms of how to successfully create and orchestrate an OSS community~\cite{bacon2012art}, and adopt the principles of open collaboration~\cite{feller2002understanding}, both on a managerial and developer level~\cite{linaaker2020public}.


Extant research reports several examples on the adoption of OSS~\cite{hollmann2013examining, DBLP:conf/oss/VenNV06, DBLP:journals/software/FitzgeraldK04}, the benefits of adopting OSS in governments~\cite{kovacs2004open,huysmans2008reasons}, and how OSS technologies can be used to restructure the public sector~\cite{hautamaki2018digital} and to develop new e-government services~\cite{kalja2007egovernment}. Studies have also been reported on the risks and critical factors related to the adoption as well as the release of OSS~\cite{kuechler2013misconceptions, scanlon2019critical, linaaker2020share}. However, research regarding the topic of how development and maintenance by PSOs are performed and organized, and how PSOs can (and should) engage with OSS projects has not received as much attention, even though highlighted as a topic in the research community since long~\cite{lundell2009panel}. One notable exception regards a study of the organization of the X-Road project (governed by the Nordic Institute for Interoperability Solutions) which has been launched by Estonia and Finland~\cite{robles2019setting}.

It should be noted that for both private organizations and PSOs which consider strategic engagement with OSS projects, which goes beyond ``ad-hoc" adoption and use of OSS, there are several challenges to address. Several strategies for how an organization may engage with different OSS projects have been conceptualized in previous research~\cite{lundell2017addressing}. Considering these strategies in a
commercial (company) context implies rather different conditions compared to a public sector context which lacks short-term commercial goals. For example, a PSO may initiate a public procurement project through which OSS from an externally provided OSS project will be adopted and deployed for internal use~\cite{shaikh2016negotiating}. Moreover, a PSO may decide to strategically engage with specific OSS projects which are of particular relevance to their organization, e.g., by providing bug reports or by seeking to establish a long-term symbiotic relationship between the own organization and an external OSS project~\cite{winter2014beyond}. A PSO may be engaged with some type of association (e.g., OS2 in Denmark\footnote{About Danish OS2 - Public Digitalization Network: \url{https://os2.eu/english}}) through which a group of PSOs join forces, in a similar way as foundations have been established for the governance of specific OSS projects, e.g., the Eclipse Foundation and The Document Foundation.



\section{Hypotheses and Research Questions}
In line with our conjecture, and informed by our previous investigation of the X-Road project~\cite{robles2019setting}, we anticipate that public sector OSS projects consist of a set of users who reside over economic, decisional, and strategic power. In contrast to a bazaar OSS project, this set of users does not (at least related to most OSS projects) actively perform any development themselves. Instead, the development is commissioned through public procurement frameworks to a set of developers (mainly contractors) who have limited power beyond the ability to provide technical input to the planning and direction of the OSS project. The developers may further consist of other stakeholders and users of the OSS project, yet they may have limited abilities to influence the planning and direction of the project. We, therefore, expect a top-down planning and communication flow from decision-makers at PSOs to developers and other users where a mix of open and closed communication channels are used.

With this background, we define a series of hypotheses in line with Mockus et al.~\cite{mockus2002two} to enable us to compare and contrast between public sector OSS projects identified through this study, and the bazaar OSS projects reported by Mockus et al~\cite{mockus2002two}. and following replications~\cite{dinh2005freebsd, ma2010developers}. It should be noted that a subset of the original hypotheses have been excluded as these focused on highlighting benefits of OSS projects compared to commercial (closed) software in terms of defect density and release pace.

\begin{itemize}
  \item[H1a] OSS developments will have a core of developers who control the code base, and will create approximately 80\% or more of the new functionality. If this core group uses only informal, ad hoc means of coordinating their work, it will be no larger than 10-15 people.
  \item[H1b] Approximately 95 \% or more of OSS developments will be performed by developers commissioned by the users of the OSS project (i.e., PSOs).
  
  \item[H2a] Projects, independent of the number of commissioned developers, coordinate their work using other mechanisms than just informal, \emph{ad hoc} arrangements. These mechanisms may include one or more of the following: explicit development processes, individual or group code ownership, and required inspections. 
  \item[H2b] Projects are planned top-down by a set of users (i.e., decision-makers at PSOs) who commission the development and communicate using open and closed communication channels.
  
  \item[H3] In successful OSS developments, a group larger by an order of magnitude than the set of users (i.e., the PSOs that commission the development) will report problems, and in other ways partake in planning activities and communication concerning the OSS project.
  
  \item[H4a] OSS developments that have a strong set of users (i.e., the PSOs that commission the development) but never achieve large numbers of general users engaged will experience limited to non-existent reuse, and a decrease in quality because of a lack of resources devoted to finding and repairing defects. 
  \item[H4b] The community of a project will primarily consist of commissioned contractors, and PSOs using, or with an interest in using the OSS.

  

\end{itemize}       

To test our hypotheses, we use the same research questions that Mockus et al.~\cite{mockus2002two}, but for our analyzed public sector OSS projects:

\begin{itemize}
    \item[RQ1] What was the process used to develop the identified public sector OSS projects?
        \begin{itemize}
            \item Addresses H1-4
        \end{itemize}
    \item[RQ2] How many people wrote code for functionality in the public sector OSS projects? How many individuals reported problems? How many individuals repaired defects? What were their affiliations and roles?
        \begin{itemize}
            \item Addresses H1, H3-4
        \end{itemize}
    \item[RQ3] Were these functions carried out by distinct groups of individuals, i.e., did individuals primarily assume a single role? Did large numbers of individuals participate somewhat equally in these activities, or did a small number of individuals do most of the work?
        \begin{itemize}
            \item Addresses H3-4
        \end{itemize}
    \item[RQ4] Where did the code contributors work in the code? Was strict code ownership enforced on a file or module level?
        \begin{itemize}
            \item Addresses H3-4
        \end{itemize}
\end{itemize}

In Section~\ref{sec:variables}, we describe how we aim to answer the research questions and address the related hypotheses.

\section{Methodology and Execution plan}
\label{sec:execution}

\begin{figure*}
  \includegraphics[width=0.95\linewidth]{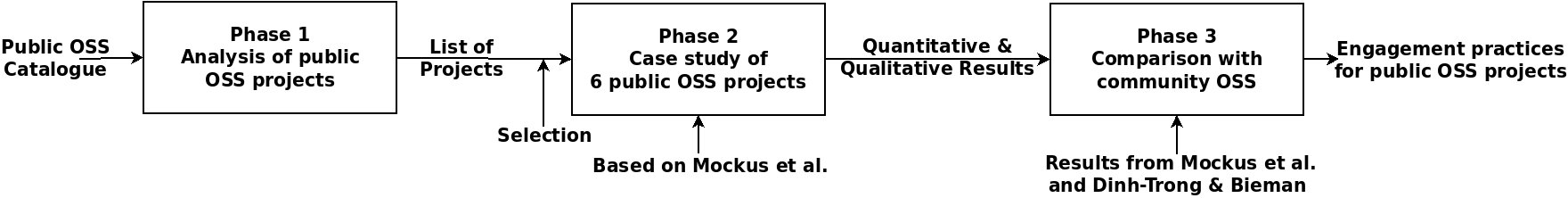}
  \caption{Execution plan in a nutshell.}
  \label{fig:execution}
\end{figure*}

Below we provide a detailed description of our research design, including the case selection criteria, underlying rationale, as well as details on our quantitative and qualitative method.

\subsection{Overall research design}
\label{research-design}

To be comparable with Mockus et al.'s findings, we plan to design our research method as close as possible to their method.
It should be noted that Mockus et al. selected two OSS projects, Apache and Mozilla, that had certain characteristics, basically that they had a large community of developers (and users). In other words, they did not randomly sample their selected OSS projects, as a vast majority of them are small in size and impact~\cite{krishnamurthy2002cave,munaiah2017curating}, but instead larger and established projects with active and vibrant developer communities (in line with the bazaar model~\cite{raymond1999cathedral, capiluppi2007cathedral}). Thus, we would first need to identify OSS projects promoted by PSOs that have a high development activity.

Hence, to pursue our research goal and address the RQs, we utilise 
purposeful sampling~\cite{patton2014qualitative} and want to report on six case studies of public sector OSS projects. The sample size is in line with Miller~\cite{miller1956magical} and motivated by the complexity that each case adds to the investigation and comparison we plan to perform. We find that the case study methodology is the most appropriate one, as Benbasat et al. consider that ``\textit{[a] case study examines a phenomenon in its natural setting, employing multiple data collection methods to gather information from a few entities. The boundaries of the phenomenon are not clearly evident at the outset of the research and no experimental control or manipulation is used}''~\cite{benbasat1987case}. So, as in Mockus et al., a mixed-methods approach will be used in this research~\cite{creswell2003advanced}, combining qualitative and quantitative data sources. This will offer a more complete perspective of the projects, based on the analysis of publicly available sources and by means of obtaining feedback from relevant stakeholders of the public sector OSS projects.

Consequently, we will divide our design in three phases (see Figure~\ref{fig:execution}): 

\begin{enumerate}
  \item In the first phase, we will identify catalogues with OSS by PSOs. These catalogues are the result of PSOs creating spaces for collaboration. In them, software created for and by PSOs at local, regional, national, and sometimes international levels are listed, and links to the source code and documentation are offered. In some cases, general-purpose software, such as LibreOffice, can also be found in those catalogues, which although widely used in PSOs does not fulfill the definition of an OSS project promoted by PSOs. We have already identified half a dozen of such catalogues, including the ones used in PSOs in Italy, France, U.S., among others. We will analyze all projects in those catalogues (which sum up to more than 20K projects), include those that are driven by a PSO, and categorize them for relevant characteristics, such as the number of collaborators (in terms of committers and bug reporters) and amount of activity (in terms of commits, bug reports and pull requests). The output of this phase will be a list of projects ordered by community size and activity. In addition, we will add information on the PSO promoting the project, such as its nationality, and level of government (international, national, regional, local), among other relevant aspects.
  
  \item In the second phase, we plan to select six projects through purposeful sampling~\cite{patton2014qualitative} out of the resulting list from the first phase. Our intention is not to choose the first six in terms of size (in number of developers and commits), but also to have a diverse set of projects (regarding nationality, level, and if developed in-house or not). These projects will be investigated quantitatively by mining relevant software repositories. We will contact representatives for projects and ask them if they are willing to provide feedback on our investigation, thereby helping to triangulate findings from our quantitative analysis. A relevant aspect for being selected to be further studied will, hence, be if they respond positively.
  
  \item In the third phase, observations from the sample of public sector OSS projects will be qualitatively compared and contrasted with the bazaar cases of Apache web server and Mozilla browser OSS projects as reported by Mockus et al.~\cite{mockus2002two}, together with the replicated cases of the  FreeBSD~\cite{dinh2005freebsd}, JBossAS, JOnAS, and Apache Geronimo projects~\cite{ma2010developers}. As a frame of reference when contrasting public sector and bazaar OSS projects, we will use the onion model~\cite{nakakoji2002evolution}, classically used to illustrate the relationship between an active and influential core of developers and a less so periphery of users in a project~\cite{crowston2006core}.
\end{enumerate}

Below, we elaborate on the methodology and its phases in further detail.

\subsection{Datasets \& Selection of Cases}
\label{subsec:datasetsAndCaseSelection}

\begin{figure}
  \includegraphics[width=0.9\linewidth]{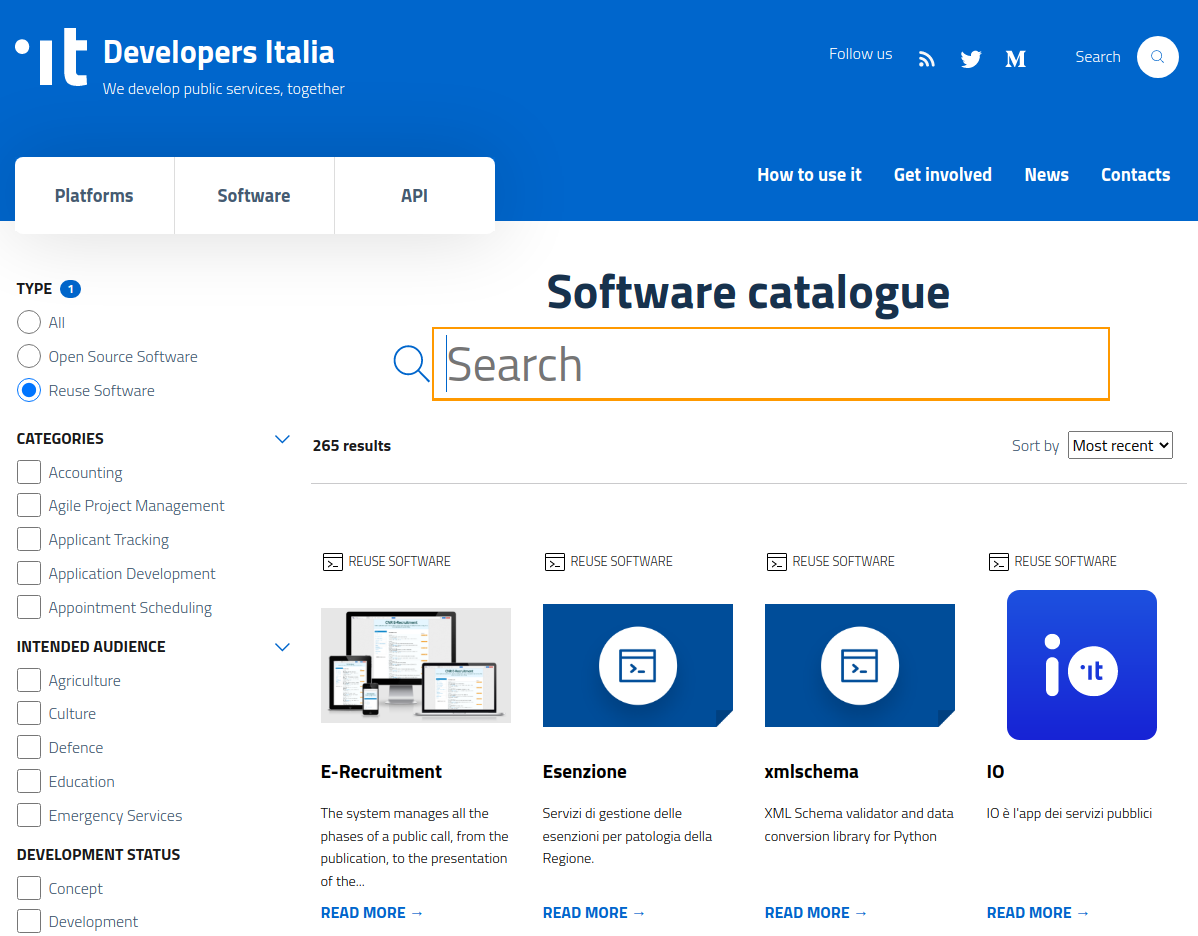}
  \caption{Snapshot of the developers.italia.it public sector OSS web catalogue from the \emph{Dipartimento per la Trasformazione Digitale} (Italian Digital Transformation Department). Provided under a Creative Commons Attribution 4.0 License, see \url{https://developers.italia.it/en/legal-notice}.}
  \label{fig:italia}
\end{figure}

We have identified at this given moment several catalogues of OSS projects for and by PSOs used in different countries. 
These include Italy\footnote{\url{https://developers.italia.it/en/software}}, France\footnote{\url{https://code.gouv.fr/}}, Canada\footnote{\url{https://code.open.canada.ca/en/index.html}}, US\footnote{\url{https://code.gov/agencies}}, Sweden\footnote{\url{https://www.offentligkod.se}}, Denmark\footnote{\url{https://os2.eu/produkter}}, Finland\footnote{\url{http://www.opencode.fi/}}, Germany\footnote{\url{}}, and the Netherlands\footnote{\url{https://developer.overheid.nl/}}.

. 
Figure~\ref{fig:italia} shows a screenshot of the Italian portal. The projects contained in these catalogues will be the input for our first phase. The catalogues were identified through personal networks and online searches, and are to be considered as a sample of potentially relevant public sector OSS projects rather than a complete data set.

The study seeks to identify six specific public sector OSS projects. For the selection process we will use four inclusion criteria (IC) for identifying these six projects:

\begin{itemize}
    \item[IC1] First, OSS that has been provided as an outcome of a software project and which has been used by at least one PSO during the time frame for the investigation (during 2023). This criteria includes software projects which have been initiated as a public procurement project irrespective of whether the provision of the OSS to the PSO (from the commercial supplier) has taken place via a public development platform for the OSS project. However, a prerequisite for the fulfillment of this criteria is that an OSS project exists on a public platform even if no representative for any PSO has been directly engaged with the development and maintenance of the software project (which may have been initiated as an OSS project on a public platform as part of a procurement contract).

    \item[IC2] Second, OSS projects provided on public development platforms for which there is some activity during the last year. This criteria seeks to exclude OSS projects for which there is inactivity on public development platforms that host the project.

    \item[IC3] Third, OSS projects for which the complete source code and related development information is publicly available which allows for inspection of the source code and for creating a running instance of the OSS from the OSS project through use of OSS licenced development tools. 
    Even if it is possible that PSOs provide source code under an OSS license for which there is a lack of OSS licensed development tools, such projects would be of limited relevance to other PSOs. 
    For this reason such projects would fail to fulfill this inclusion criterion.

    \item[IC4] Fourth, OSS projects for which commercial support and services related to the project may be obtained (and have been obtained) by PSOs. This criteria presupposes that there exists at least one PSO which has obtained a commercial contract with service providers related to the OSS project.
\end{itemize}

IC1-2 will be applied through mining process described in the first phase of our study, while IC3-4 will be manually verified in the second phase when identifying the six projects to be investigated (see Figure~\ref{fig:execution}).

The output of the selection process will be a set of projects that have certain characteristics (as discussed above) and which we will analyze more in-depth. We need therefore access to their source code repository (nowadays, usually a git repository) and to the bug tracking system. In most cases, from an initial inspection of the identified catalogues, we have seen that this is a GitHub or a GitLab repository. In particular, with this set of projects, we will follow the research design described in the next subsections.

\subsection{Variables}
\label{sec:variables}

Below, we list the variables that we will analyze to address the RQs. We aim to follow the same method (and thus collect the same variables) as in Mockus et al.~\cite{mockus2002two}, to make the comparison in the third phase possible (see Figure~\ref{fig:execution}).

\subsubsection{RQ1: Development process}

To produce an accurate description of the OSS development processes, one of the authors of Mockus et al. wrote a draft description of the process for each project (i.e., Apache and Mozilla), then had it reviewed by members of the core OSS development teams. The scope of the description includes information on:

\begin{itemize}
  \item Roles and responsibilities
  \item Identifying work to be done
  \item Assigning and performing development work 
  \item Pre-release testing
  \item Inspections
  \item Managing releases
\end{itemize}

We plan to replicate this process by writing a draft on our own based on evidence found on the project's web page and repositories and asking members of the project to review and validate it. The interviews will also allow for follow-up questions to clarify and explore the development process of the OSS projects. 

\subsubsection{RQ2: Community size}

To address this question, as in Mockus et al., we will identify those who have submitted code (discriminating between those who add functionality and those who fix bugs) and those who file bug reports to the bug tracking system.

\subsubsection{RQ3: Involvement \& roles}

Mockus et al. plotted the cumulative proportion of code changes and bug reports against the number of contributors and accounted for the share of contributions of the top 15 developers (what they called the \emph{core group}, following the \emph{onion model}). We will also examine whether the size of commits by the \emph{most active developers} is statistically larger than those done by the rest, and how the participation is in terms of adding new functionalities. We will measure as well the tenure of \emph{most active developers} and the rest. 

For bug reports, we will investigate the distribution of those who have reported the bugs, and how much of the total share has been done by the \emph{most active developers}. 

\subsubsection{RQ4: Code ownership}

To address this question, we will track those who have made modifications to files and see if only one developer is in charge of them or if code ownership is shared among more developers.





\subsection{Qualitative analysis}
\label{subsec:qualitativeanalysis}

As highlighted in Section~\ref{research-design}, we aim to validate our investigation through interviews with individuals within the communities of the respective cases. This regards both the process description as highlighted by RQ1, but also our general findings in terms of RQ2-4. Individuals will be sampled based on their social and technical activity, and their (potential) organizational affiliation. A minimum of two individuals will be sampled per project, with additional ones sampled if inconsistency among their responses and reflections would occur.

Interviews will be conducted by two of the authors, with one leading the interview and the second taking notes. Observations and findings will be presented to the interviewee who will then be encouraged to openly reflect and respond based on their own experiences. Interviews will be recorded and transcribed, and structurally coded~\cite{saldana2021coding} based on the respective RQs. The coded data will then be used to triangulate our previous findings and used in our analysis of each project. The data will further provide input to our synthesis and comparison between the bazaar projects investigated by Mockus et al.~\cite{mockus2002two} and replicated works~\cite{dinh2005freebsd, ma2010developers}, and the public sector OSS projects investigated in this study.

The authors will abide by the ethical guidelines provided by the Swedish Research Council~\cite{vr2017god}. In terms of data management specifically, interviewees' identities and affiliations will be anonymized in transcriptions and any reporting based on the data. Each interviewee will be provided with a copy of the transcript and encouraged to address any misunderstanding or mentions of sensitive information. Any quotes considered for use in reporting will be verified with the concerned interviewees. All recordings will be destroyed after the reporting of this study. 

\section{Threats to validity}

As this is a replication study, the method used must be as aligned as possible with that of the original study~\cite{mockus2002two} for results to be comparable. All six authors have therefore independently reviewed and compared the methodology of the current and the original studies, and together discussed any misalignment that could be identified.

Certain differences in design decisions do however apply, e.g., in terms of case sampling. Beyond the fact that the two cases of the original study~\cite{mockus2002two} had large and vibrant communities, the authors chose their cases also because they had personal experience and rootings in their respective communities. In our study, we are specifically interested in characterizing public sector OSS projects, and how these may differ from bazaar OSS projects, such as those investigated by the original study. To avoid any bias in our sampling, we will mine software catalogues listing OSS projects used and developed by PSOs in different countries. Six different types of projects will then be sampled based on our defined inclusion criteria.

As the sample is limited, we do not claim that the results are generalizable. Rather, we provide a contextual description of what we refer to as public sector OSS projects and exemplify how their development model (may) differ from that of bazaar OSS projects in general and the commonly used and associated onion model. Through both our qualitative and quantitative investigations of the projects, we will provide in-depth characteristics of the six projects to enable anecdotal generalization to cases with similar characteristics. Further generalization is an open topic for future work.

In contrast to the original study, we do not expect to have in-depth knowledge of the six cases to be sampled. Hence, to limit the risk of research bias and misinterpretations of observations we will allow community representatives to review and criticize our findings (RQ1 specifically, and RQ2-6 in general).

Another threat regards the construct validity of our study - if what we consider as bazaar projects instantiated in the cases of the Apache web server and Mozilla browser OSS projects is still valid today. The report by Mockus et al.~\cite{mockus2002two} dates from 2002, and practices and processes may have evolved, why it may be considered relevant to revisit the two communities and how they currently work. However, as this is a replication study, we are limited to using the reporting by Mockus et al.~\cite{mockus2002two} as our baseline. Further, we note that the study by Mockus et al.~\cite{mockus2002two} has been replicated and generalized by several studies through the years (see~\cite{dinh2005freebsd, ma2010developers}) since its reporting, why we still consider it a valid representation of what we consider as a bazaar project. Further, we will include these replications of Mockus et al.~\cite{mockus2002two} into the comparison between the public sector and bazaar OSS projects to get further nuance of what characterizes a bazaar project.

\section{Implications}

The public sector has lagged behind the adoption of OSS and the evolution in terms of maturity that the industry has undergone in the past twenty years. Yet, there are several opportunities identified in terms of leveraging OSS as a means of creating economic growth~\cite{greenstein2014digital, nagle2019government, blind2021impact}, and also enabling policy objectives, e.g., related to interoperability~\cite{lundell2017addressing}, transparency~\cite{ofe2022strategic}, and digital sovereignty~\cite{nagle2022strengthening}. Hence, PSOs will need to consider how these objectives may be achieved. 

PSOs could, in this regard, benefit from lessons learned by industry and established best practices. There are, however, many specific conditions under which all PSOs operate that are rather different from
companies. Thus, any (top-down) policy with ``\textit{recipes}'' taken from company contexts may be a very good reference point, although these need to be grounded in the conditions under which PSOs operate. Given the special circumstances of PSOs, e.g., in terms of missing technical capabilities, and the need to abide by public procurement regulations and procurement practices, these lessons and practices need to be scrutinized in public sector contexts and redesigned accordingly.

Based on the six case studies of public sector OSS projects, and the comparison with the earlier reported cases of bazaar OSS projects, we aim to design engagement practices that can fill this need. Our ambition is that these practices will guide PSOs in how to initiate, and engage in public sector OSS projects, in a way that enables the sought-for benefits. A better understanding of how PSOs might best leverage OSS could lead to better government services, and ultimately improve the quality of life for citizens.



%
%

\bibliographystyle{ACM-Reference-Format}
\bibliography{software}

\end{document}

%% file: macros.tex
\newcommand{\nb}[2]{
	{
		{\color{black}{
				\fbox{\bfseries\sffamily\scriptsize#1}
				{\sffamily$\triangleright~${\it\sffamily #2}$~\triangleleft$}
	}}}
}

\ifdraft
\newcommand\johan[1]{\nb{Johan}{\color{blue}#1}}
\newcommand\grex[1]{\nb{Gregorio}{\color{red}#1}}
\newcommand\paco[1]{\nb{Francisco}{\color{teal}#1}}
\newcommand\jonas[1]{\nb{Jonas}{\color{violet}#1}}
\newcommand\bjoern[1]{\nb{Björn}{\color{purple}#1}}
\newcommand\sachiko[1]{\nb{Sachiko}{\color{green}#1}}
\newcommand\reviewer[1]{\nb{Reviewer}{\color{red}#1}}
\newcommand{\fixme}[1]{{\textcolor{red}{[FIXME] #1}}\xspace}

\else
\usepackage[disable]{todonotes}
\newcommand\johan[1]{}
\newcommand\grex[1]{}
\newcommand\paco[1]{}
\newcommand\jonas[1]{}
\newcommand\bjoern[1]{}
\newcommand\sachiko[1]{}
\newcommand\reviewer[1]{}
\newcommand{\fixme}[1]{}

\fi

\usepackage[inline]{enumitem}

